\documentclass[preprint,amsmath,amssymb,11pt,english,aps,showkeys,showpacs]{revtex4}
\usepackage[english]{babel}
\usepackage{graphicx}
\usepackage{graphics}
\usepackage{amsmath}
\usepackage{dcolumn}
\usepackage{amssymb}
\usepackage{bm}
\usepackage{hyperref}
\begin{document}
\title{Effects of the Cornell-type potential on a position-dependent mass system in Kaluza-Klein theory}

\author{E. V. B. Leite}
\email{ericovbleite@gmail.com}
\affiliation{Departamento de F\'isica e Qu\'imica, Universidade Federal do Esp\'irito Santo, Av. Fernando Ferrari, 514, Goiabeiras, 29060-900, Vit\'oria, ES, Brazil.}

\author{H. Belich}
\email{belichjr@gmail.com}
\affiliation{Departamento de F\'isica e Qu\'imica, Universidade Federal do Esp\'irito Santo, Av. Fernando Ferrari, 514, Goiabeiras, 29060-900, Vit\'oria, ES, Brazil.}

\author{R. L. L. Vit\'oria}
\email{ricardo.vitoria@ufes.br/ricardo.vitoria@pq.cnpq.br}
\affiliation{Departamento de F\'isica e Qu\'imica, Universidade Federal do Esp\'irito Santo, Av. Fernando Ferrari, 514, Goiabeiras, 29060-900, Vit\'oria, ES, Brazil.}

\begin{abstract}
In this paper, we have investigated a scalar particle with position-dependent mass subject to a uniform magnetic field and a quantum flux, both coming from the background which is governed by the Kaluza-Klein theory. By modifying the mass term of the scalar particle, we insert the Cornell-type potential. In the search for solutions of bound states we determine the relativistic energy profile of the system in this background of extra dimension. Particular cases of this system are analyzed and a quantum effect can be observed: the dependence of the magnetic field on the quantum numbers of the solutions.
\end{abstract}

\keywords{Kaluza-Klein theory, Landau quantization, Aharanov-Bohm effect for bound states, central potentials}
\pacs{03.65.Vf, 11.30.Qc, 11.30.Cp}

\maketitle

\section{Introduction}\label{sec1}

In quantum mechanics systems, we consider the mass of the particle that is immersed in the system as a constant, for example, the effective mass of the hydrogen atom problem \cite{eisberg}, the mass of the three-dimensional harmonic oscillator \cite{eisberg} and the mass of an electrically charged particle subject to a perpendicular uniform magnetic field to the plane of motion of the particle (Landau quantization) \cite{landau}. These are classic problems and idealized prototypes which provide results that can be approximated with real problems and with large applications. However, recently, several studies have emerged with the proposition that effective mass can be a function of the position \cite{smp6, smp7, smp8, smp9, smp10, smp11, smp12, smp13, smp14, smp15, smp16, smp17, smp18, smp19, smp20, smp21}. The justification for this change in effective mass is based on systems where there are several applications, for example, semiconductor heterostructures \cite{smp}, electronic properties of the semiconductors \cite{bastard}, quantum wells, wires and dots \cite{harr, smp1, smp2, smp3}, quantum liquids \cite{smp4} and$ \ ^{3}$He clusters \cite{smp5}. It is noteworthy that quantum systems where mass is a function of the position is known in the literature as a position-dependent mass quantum system \cite{smp22, smp23, smp24}.

Position-dependent mass quantum systems have been investigated in the relativistic context, for example, the pionic atom \cite{greiner}, in solution of the Dirac equation \cite{smpr}, in implications in atomic physics \cite{smpr1}, in the quark-antiquark interaction \cite{smpr2}, in effects of external fields on a two-dimensional Klein Gordon particle under pseudo-harmonic oscillator interaction \cite{smpr3}, in noncommutative space \cite{smpr4}, in the cosmic spacetime \cite{eug, smpr5}, in the global monopole spacetime \cite{eug1}, in the rotating cosmic string spacetime \cite{smpr6}, in the spacetime with torsion \cite{me, me1, me2}, in possible scenarios of Lorentz symmetry violation \cite{bb, me3, me4}, on the Klein-Gordon oscillator \cite{bf, me5, me6, me7}, on the Majorana fermion \cite{rb}, in the Som-Raychaudhuri spacetime \cite{smpr7, me8} and in Kaluza-Klein theory \cite{smpr8}.

The first attempt in explain the physics interaction (gravitation and electromagnetism) in a unified way is the propose of the Kaluza-Klein theory (KKT) \cite{kaluza, klein}. They established that the electromagnetism can be introduced through an extra (compactified) dimension in the spacetime, where the spatial dimension becomes five-dimensional.

Recently, the KKT has been the background for quantum systems research. In the non-relativistic context, we have studies in global effects due to cosmic defects \cite{kk}, in Aharonov-Bohm effect for bound states \cite{kk1} and in Landau levels \cite{kk2}; in the relativistic context, we have studies in Aharonov-Bohm effect for bound states on the confinement of a scalar particle to a Coulomb-type potential \cite{smpr8}, in loop variables for a class of spacetimes, in geometric phases in graphene \cite{kk4}, on the Klein-Gordon oscillator \cite{kk5} and in the Dirac field \cite{kk6}. However, one point that has not yet been investigated is the interaction between a position-dependent mass scalar field with a linear potential plus a Coulomb-type potential in an environment described by the KKT theory.

In this paper, we investigate the effects of central potentials on a position-dependent mass scalar field in a spacetime of (1+4) dimensions described by KKT, where we chose a particular case for the gauge field with extra dimension which characterizes the relativistic Landau gauge \cite{eug, me1} and describes the presence of a magnetic quantum flux along the $z$-axis providing an effect analogous to the Aharanov-Bohm effect for bound states \cite{ab, ab1}. Given this, we solve the Klein-Gordon equation in this background, where, analytically, we obtain solutions of bound states for a scalar field subject to a linear central potential plus a Coulomb-type potential, that is, a Cornell-type potential. We then detail these solutions of bound states for only the linear potential and only the Coulomb-type potential, where in all cases the influence of the gauge configuration from the extra dimension on the relativistic energy levels of the system is perceptible.

The structure of this paper is as follows: in the Sec. (\ref{sec2}), we investigate a position-dependent mass scalar field in a background described by KKT, where, in our first analysis (\ref{sec2a}) we describe the relativistic quantum dynamics of this position-dependent mass system subject to Cornell-type interaction, of which, analytically, we determine its energy profile. Then, we analyze particular cases of this system, (\ref{sec2b}) where we consider only the Coulomb-type central potential and (\ref{sec2c}) only the central linear potential; in the Sec. (\ref{sec3}), we present our conclusions.

\section{Effects of central potentials on a position-dependent mass system in a Kaluza-Klein theory}\label{sec2}

The main idea behind the KKT \cite{kaluza, klein} is that the spacetime is five-dimensional with the purpose of unifying electromagnetism and gravitation. In this way, we can work with general relativity in five dimensions. The information about the electromagnetism is given by introducing a gauge field $A_{\mu}(x)$ in the line element of the spacetime as \cite{smpr8, kk2} $(c=\hbar=1)$:
\begin{eqnarray}\label{eq01}
ds^{2}=-dt^2+d\rho^2+\rho^2d\varphi^2+dz^2+\left[dw+KA_{\mu}(x)d\varphi\right]^2,
\end{eqnarray}
where $\rho=(x^2+y^2)^{1/2}$ is radial coordinate, $0\leq\varphi\leq 2\pi$, $-\infty<z<\infty$, $K$ is the Kaluza constant \cite{kk} and $w$ is the extra dimension. We are interested in investigated a massive scalar field of position-dependent mass subject to a uniform magnetic field and under the Aharonov-Bohm effect for bound states \cite{ab, ab1}. Then, based on Refs. \cite{smpr8, kk5}, we can introduce a uniform magnetic field $B_0$ and a quantum flux $\Phi$ through the line element of the Minkowski spacetime (\ref{eq01}) in the form
\begin{eqnarray}\label{eq02}
ds^{2}=-dt^2+d\rho^2+\rho^2d\varphi^2+dz^2+\left[dw+\left(\frac{B_0\rho^2}{2}+\frac{\Phi}{2\pi}\right)d\varphi\right]^2,
\end{eqnarray}
where the gauge field is given by the component
\begin{eqnarray}\label{eq03}
A_{\varphi}=\frac{B_0\rho^2}{2K}+\frac{\Phi}{2\pi K},
\end{eqnarray}
which gives rise to a uniform magnetic field $\vec{B}=\nabla\times\vec{A}=K^{-1}B_0\hat{z}$ \cite{kk2}, where $\hat{z}$ is unitary vector in the $z$-direction. Therefore, for a position-dependent mass scalar field in this five-dimensional spacetime, the Klein-Gordon equation is written in the form \cite{smpr8}:
\begin{eqnarray}\label{eq04}
\partial_{\mu}(g^{\mu\nu}\sqrt{-g}\partial_{\nu})\phi-\sqrt{-g}[m+S(\vec{r})]^2\phi=0,
\end{eqnarray}
where $g^{\mu\nu}$, with $\mu,\nu=0,1,2,3,4$, is inverse metric tensor, $m$ is rest mass of the scalar field and $S(\vec{r})=S(\rho)$ is scalar central potential. In this way, from Eq. (\ref{eq02}), the Eq. (\ref{eq04}) becomes
\begin{eqnarray}\label{eq05}
-\frac{\partial^2\phi}{\partial t^2}&+&\frac{\partial^2\phi}{\partial\rho^2}+\frac{1}{\rho}\frac{\partial\phi}{\partial\rho}
+\frac{1}{\rho^2}\frac{\partial^2\phi}{\partial\varphi^2}-\left(\frac{\Phi}{\pi\rho^2}-B_0\right)\frac{\partial^2\phi}{\partial\varphi\partial w}+\frac{\partial^2\phi}{\partial w^2}
+\frac{\partial^2\phi}{\partial z^2}+\frac{\Phi^2}{4\pi^2\rho^2}\frac{\partial^2\phi}{\partial w^2} \nonumber \\
&-&\frac{B_0\Phi}{2\pi}\frac{\partial^2\phi}{\partial w^2}
+\frac{B_0^2\rho^2}{4}\frac{\partial^2\phi}{\partial w^2}-[m+S(\rho)]^2=0,
\end{eqnarray}
which represents the interaction of a position-dependent mass scalar field with a uniform magnetic field in the five-dimensional spacetime described by a KKT.

\subsection{Cornell-type potential}\label{sec2a}

The Cornell potential, which consists of a linear potential plus a Coulomb potential, is a particular case of the quark-antiquark interaction, which has one more harmonic type term \cite{smpr2}. The Coulomb potential is responsible by the interaction at small distances and the linear potential leads to the confinement. Recently, the Cornell potential has been studied in the ground state of three quarks \cite{fn}. However, this type of potential is worked on spherical symmetry; in cylindrical symmetry, which is our case, this type of potential is known as Cornell-type potential \cite{smpr7}. This type of interaction has been studied in the Refs. \cite{eug, smpr6, me2, me4, me5, me7, smpr7}. Given this, let us consider the central
\begin{eqnarray}\label{eq06}
S(\rho)=\frac{a}{\rho}+b\rho,
\end{eqnarray}
where $a$ and $b$ are constants. In addition, the solution to Eq. (\ref{eq05}) can be written in the form
\begin{eqnarray}\label{eq07}
\phi(\rho,\varphi,z,w,t)=e^{-i(\mathcal{E}t-l\varphi-kz-qw)}u(\rho),
\end{eqnarray}
where $l=0,\pm1,\pm2,\ldots$, $-\infty<k<\infty$, $q$ is a constant and $u(\rho)$ is a radial wave function. By substituting the Eqs. (\ref{eq06}) and (\ref{eq07}) into the Eq. (\ref{eq05}), we obtain
\begin{eqnarray}\label{eq08}
\frac{d^2u}{d\rho^2}+\frac{1}{\rho}\frac{du}{d\rho}-\frac{\iota^2}{\rho^2}u-\frac{2am}{\rho}u-2bm\rho u-\Omega^2\rho^2u+\alpha u=0,
\end{eqnarray}
where we define the parameters
\begin{eqnarray}\label{eq09}
\alpha=\mathcal{E}^2-m^2-k^2-q^2-m\omega\left(l-\frac{q\Phi}{2\pi}\right)-2ab; \ \ \ \iota^2=\left(l-\frac{q\Phi}{2\pi}\right)^2; \ \ \ \Omega^2=b^2+\frac{m^2\omega^2}{4},
\end{eqnarray}
with $\omega=\frac{qB_0}{m}$.

Let us define $\varrho=\sqrt{\Omega}\rho$, then the Eq. (\ref{eq08}) becomes
\begin{eqnarray}\label{eq10}
\frac{d^2u}{d\varrho^2}+\frac{1}{\varrho}\frac{du}{d\varrho}-\frac{\iota^2}{\varrho^2}u-\frac{\beta}{\varrho}u-\delta\varrho u-\varrho^2u+\frac{\alpha}{\Omega}u=0,
\end{eqnarray}
where
\begin{eqnarray}\label{eq11}
\beta=\frac{2am}{\sqrt{\Omega}}; \ \ \ \delta=\frac{2bm}{\Omega^{3/2}}.
\end{eqnarray}

The radial wave function $u(\varrho)$ must be well behaved at $\varrho\rightarrow0$ and at $\varrho\rightarrow\infty$. Then, the analysis of the asymptotic behavior of the Eq. (\ref{eq10}) at $\varrho\rightarrow0$ and at $\varrho\rightarrow\infty$ gives us the following solution in terms of an unknown function $H(\varrho)$ \cite{eug, eug1}:
\begin{eqnarray}\label{eq12}
u(\varrho)=\varrho^{|\iota|}e^{-\frac{1}{2}\varrho(\varrho+\delta)}H(\varrho).
\end{eqnarray}

Then, by substituting the Eq. (\ref{eq12}) into the Eq. (\ref{eq10}), we have
\begin{eqnarray}\label{eq13}
\frac{d^2H}{d\varrho^2}+\left[\frac{2|\iota|+1}{\varrho}-2\varrho-\delta\right]\frac{dH}{d\varrho}+
\left[\frac{\alpha}{\Omega}+\frac{\delta^2}{4}-2-2|\iota|-\frac{\delta}{2\varrho}(2|\iota|+1)-\frac{\beta}{\varrho}\right]H=0,
\end{eqnarray}
which is known in the literature as biconfluent Heun equation \cite{eug, heun} and $H(\varrho)$ is a biconfluent Heun function: $H(\varrho)=H_{b}\left(2|\iota|,\delta,\frac{\alpha}{\Omega}+\frac{\delta^2}{4},2\beta;\varrho\right)$.

The biconfluent Heun equation has two singular points, where one is the origin and the other is infinity \cite{eug}. The origin is a regular singular point. Given this, the Eq. (\ref{eq13}) has at least one solution around the origin given by the power series \cite{arf}:
\begin{eqnarray}\label{eq14}
H(\varrho)=\sum_{j=0}^{\infty}c_{j}\varrho^{j}.
\end{eqnarray}

By substituting the Eq. (\ref{eq14}) into the Eq. (\ref{eq13}), we obtain the recurrence relation of the biconfluent Heun series
\begin{eqnarray}\label{eq15}
c_{j+2}=\frac{[\tau+\delta(j+1)]c_{j+1}-(\theta-2j)c_j}{(j+2)(j+2+2|\iota|)}
\end{eqnarray}
and the coefficients
\begin{eqnarray}\label{eq16}
c_{1}=\frac{\tau}{1+2|\iota|}; \ \ \ c_{2}=\frac{(\tau+\delta)c_{1}-\theta c_0}{2(2+2|\iota|)}=\frac{c_0}{4(1+|\iota|)}\left[\frac{(\tau+\delta)\tau}{1+2|\iota|}-\theta\right],
\end{eqnarray}
where we define the new parameters
\begin{eqnarray}\label{eq17}
\theta=\frac{\alpha}{\Omega}+\frac{\delta^2}{4}-2-2|\iota|; \ \ \ \tau=\frac{\delta}{2}(2|\iota|+1)+\beta.
\end{eqnarray}

As we are interested in solutions of bound states, therefore, to obtain a finite degree polynomial for the biconfluent Heun series, we must truncate the power series, and this is possible through the following conditions \cite{eug, me1}:
\begin{eqnarray}\label{eq18}
c_{n+1}=0; \ \ \ \theta=2n
\end{eqnarray}
where $n=1,2,3,\ldots$ represents the radial modes. In order to analyze these conditions, we must assign values to $n$. In this case, consider $n=1$, that is, the radial mode corresponding to the lowest energy state of the system. Therefore, the condition $c_{n+1}=0$ produces $c_2=0$, which from Eq. (\ref{eq16}) we obtain
\begin{eqnarray}\label{eq19}
\Omega_{l,1}^3-\frac{2a^2m^2}{1+2|\iota|}\Omega_{l,1}^2-\frac{4abm^2(1+|\iota|)}{1+2|\iota|}\Omega_{l,1}-\frac{b^2m^2(2|\iota|+3)}{2}=0,
\end{eqnarray}
where we choose the frequency (or the magnetic field) as the parameter of adjustment of the condition $c_{n+1}=0$, not only for the radial mode $n=1$, but for any value of $n$. In addition, as the parameter $\Omega$ depends on the magnetic field as established in the Eq. (\ref{eq20}), where we have simplified our notation by labelling:
\begin{eqnarray}\label{eq20}
\omega_{l,1}=\frac{2}{m}\sqrt{\Omega_{l,1}-b^2} \ \leftrightarrow \ B_0^{l,1}=\frac{2}{q}\sqrt{\Omega_{l,1}-b^2}.
\end{eqnarray}
It is noteworthy that a third-degree algebraic equation has at least one real solution and it is exactly this solution that gives us the allowed values of the magnetic field for the lowest state of the system, which we do not write it because its expression is very long. We can note, from the Eq. (\ref{eq19}), that the possible values of the magnetic field depend on the quantum numbers of the system and the parameters associated with the background governed by a KKT. In addition, for each relativistic energy level, we have a different relation of the magnetic field to the parameters of the gauge field given by KKT, of the parameters associated to the Cornell-type potential and the quantum numbers of the system $\{l,n\}$. For this reason, we have labelled the parameters $\Omega$, $\omega$ and $B_0$ in the Eqs. (\ref{eq19}) and (\ref{eq20}).

For our analysis to become complete we must take $n=1$ in the condition $\theta=2n$, that is, $\theta=2$, which gives us the expression
\begin{eqnarray}\label{eq21}
\mathcal{E}_{k,l,1}=\pm\sqrt{m^2+k^2+q^2+2ab+m\omega_{l,1}\left(l-\frac{q\Phi}{2\pi}\right)+2\Omega_{l,1}\left(2+\left|l-\frac{q\Phi}{2\pi}\right|\right)
-\frac{b^2m^2}{\Omega^2_{l,1}}}
\end{eqnarray}
Then, by substituting the real solution of Eq. (\ref{eq19}) into the Eq. (\ref{eq21}) it is possible to obtain the allowed values of the relativistic energy for the radial mode $n=1$ of a position dependent mass scalar particle in a background governed by the metric given in Eq. (\ref{eq02}). In contrast to Refs. \cite{kk, kk5}, we can see that the lowest energy state defined by the real solution of algebraic equation given in the Eq. (\ref{eq19}) plus the expression given in the Eq. (\ref{eq21}), is defined by the radial mode $n=1$, instead of $n=0$. This effect arises due to the presence of the Cornell-type central potential in the system. Note that it is necessary physically that the lowest energy state is $n=1$ and not $n=0$, otherwise the opposite would imply that $c_1=0$, which requires that the rest mass of the scalar particle be zero, that is contrary to the proposal of this investigation.

We can specify our analysis through the Cornell-type potential, that is, by taking the parameters $a\rightarrow0$ or $b\rightarrow0$, imposing that interaction be produced only by the linear potential or the Coulomb-type potential, respectively. We can observe that if we take the limit $a\rightarrow0$ and $b\rightarrow0$ in Eq. (\ref{eq06}), hence, we recover the Klein-Gordon equation without position-dependent mass in a KKT background defined by the gauge configuration given in Eq. (\ref{eq03}). Therefore, we would have in Eq. (\ref{eq05}) the Klein-Gordon equation for the relativistic Landau quantization and a quantum flux. In this case, the solution to the radial wave equation (\ref{eq08}) would be given in terms of the confluent hypergeometric function. This analysis has been made in Refs. \cite{kk, kk5}, where the magnetic field doesn't have restricted values.

\subsection{Coulomb-type potential}\label{sec2b}

In this particular case, it means that $b\rightarrow0$. Thus, the scalar central potential given in Eq. (\ref{eq06}) is rewritten as:
\begin{eqnarray}\label{eq22}
S(\rho)=\frac{a}{\rho}.
\end{eqnarray}

The Eq. (\ref{eq22}) represents a Coulomb-type potential. This type of potential has been studied in propagation of gravitational waves \cite{pc} and quark models \cite{pc1}. There are also investigations of the Coulomb-type potential in condensed matter systems, an atom with electric quadrupole moment \cite{pc2}  and magnetic quadrupole moment \cite{pc3}, neutral particle with permanent magnetic dipole moment \cite{pc4}, in molecules \cite{pc5, pc6, pc7} and in pseudo-harmonic interactions \cite{pc8, pc9}. Therefore, in this particular case, the Eq. (\ref{eq19}) is reduced and gives us the allowed values of the magnetic field for the radial mode $n=1$:
\begin{eqnarray}\label{eq23}
\omega_{l,1}=\frac{4a^2m}{\left[1+2\sqrt{\left(l-\frac{q\Phi}{2\pi}\right)^2+a^2}\right]} \ \leftrightarrow \ B_0^{l,1}=\frac{4a^2m^2}{q\left[1+2\sqrt{\left(l-\frac{q\Phi}{2\pi}\right)^2+a^2}\right]}.
\end{eqnarray}

We can note that the allowed values of the magnetic field are connected by the quantum flux $\Phi$ through the shift in the eigenvalues of the angular momentum, $l_{\text{eff}}=\left(l-\frac{q\Phi}{2\pi}\right)$, that is, an effect analogous to the Aharonov-Bohm effect for bound states \cite{ab, ab1}, making them a periodic function with periodicity $\Phi_0=\pm\frac{2\pi}{q}\nu$, with $\nu=0,1,2,\ldots$, that is, $B_{0}^{l,1}(\Phi+\Phi_0)=B_{0}^{l\mp\nu,1}(\Phi)$.

The Eq. (\ref{eq21}) is also reduced as follows:
\begin{eqnarray}\label{eq24}
\mathcal{E}^{2}_{k,l,1}=m^2+k^2+q^2+m\omega_{l,1}\left[2+\sqrt{\left(l-\frac{q\Phi}{2\pi}\right)^2+a^2}+l-\frac{q\Phi}{2\pi}\right].
\end{eqnarray}

Then, by substituting the Eq. (\ref{eq23}) into the Eq. (\ref{eq24}), we obtain
\begin{eqnarray}\label{eq25}
\mathcal{E}_{k,l,1}=\pm\sqrt{m^2+k^2+q^2+\frac{4a^2m^2}{[1+2\sqrt{(l-q\Phi/2\pi)^2+a^2}]}
\left(2+\sqrt{\left(l-\frac{q\Phi}{2\pi}\right)^2+a^2}+l-\frac{q\Phi}{2\pi}\right)}.
\end{eqnarray}

The Eq. (\ref{eq25}) corresponds to the allowed values of relativistic energy for the radial mode $n=1$ of a position-dependent mass particle subject to the Coulomb-type potential in a possible scenario described by a KKT. By comparing the Eq. (\ref{eq25}) with the result obtained in Refs. \cite{kk, kk5}, we can note that the presence of the Coulomb-type potential modifies the energy profile of the system. This modification can be explained by the radical breaking of degeneracy of the Landau levels and by the representation of the lowest energy state that is defined by the radial mode $n=1$, instead of $n=0$. By comparing the Eq. (\ref{eq25}) with the result obtained in Ref. \cite{smpr8}, we can also assume that the presence of a uniform magnetic field modifies the energy profile of the system. In addition, the allowed values of relativistic energy (\ref{eq25}) are influenced by the quantum flux $\Phi$ through the shift in the eigenvalues of the angular momentum $l_{\text{eff}}=\left(l-\frac{q\Phi}{2\pi}\right)$, that is, an effect analogous to the Aharonov-Bohm effect for bound states \cite{ab, ab1}, making them a periodic function with periodicity $\Phi_0=\pm\frac{2\pi}{q}\nu$, with $\nu=0,1,2,\ldots$, that is, $\mathcal{E}_{l,1}(\Phi+\Phi_0)=\mathcal{E}_{l\mp\nu,1}(\Phi)$. This latter characteristic may be of interest in persistent current calculations \cite{me, smpr8, cp}. The result given in Eq. (\ref{eq25}), less than the term $q^2$ that stems from the KKT, is analogous to the result obtained in Ref. \cite{me2}. However, it is noteworthy that, in the latter case, the Landau gauge is inserted into the Klein-Gordon equation by the minimum coupling.

\subsection{Linear potential}\label{sec2c}

Now, let us consider the particular case $a\rightarrow0$. Thus, the scalar central potential given in Eq. (\ref{eq06}) is rewritten as:
\begin{eqnarray}\label{eq26}
S(\rho)=b\rho.
\end{eqnarray}

The Eq. (\ref{eq22}) represents a linear central potential. There are studies involving the linear potential in atomic and molecular physics \cite{pl, pl1, pl2, pl3, pl4}, quantum bouncer \cite{pl5, pl6}, motion of a quantum particle in a uniform force field \cite{landau, ball} and in relativistic quantum systems \cite{eug, eug1, me, me1, me4, me6, rb, me8}. Therefore, in this particular case, the Eq. (\ref{eq19}) is reduced and gives us the allowed values of the magnetic field for the radial mode $n=1$:
\begin{eqnarray}\label{eq27}
B_0^{l,1}=\frac{2}{q}\sqrt{\left[\frac{b^2m^2}{2}\left(2\left|l-\frac{q\Phi}{2\pi}\right|+3\right)\right]^{2/3}-b^2}.
\end{eqnarray}

Again, we can observe that the allowed values of the magnetic field for the radial mode $n=1$ (\ref{eq27}) are influenced by the quantum flux $\Phi$ through the shift in the eigenvalues of the angular momentum $l_{\text{eff}}=\left(l-\frac{q\Phi}{2\pi}\right)$, that is, an effect analogous to the Aharonov-Bohm effect for bound states \cite{ab, ab1}, making them a periodic function with periodicity $\Phi_0=\pm\frac{2\pi}{q}\nu$, with $\nu=0,1,2,\ldots$, that is, $B_{0}^{l,1}(\Phi+\Phi_0)=B_{0}^{l\mp\nu,1}(\Phi)$.

The Eq. (\ref{eq21}) is also reduced as follows:
\begin{eqnarray}\label{eq28}
\mathcal{E}^2_{k,l,1}=m^2+k^2+q^2+m\omega_{l,1}\left(l-\frac{q\Phi}{2\pi}\right)+2\Omega_{l,1}\left(2+\left|l-\frac{q\Phi}{2\pi}\right|\right)
-\frac{b^2m^2}{\Omega^2_{l,1}}.
\end{eqnarray}

By substituting the Eq. (\ref{eq27}) into the Eq. (\ref{eq28}), we obtain
\begin{eqnarray}\label{eq29}
\mathcal{E}_{k,l,1}&=&\pm\Bigg[m^2+k^2+q^2+
2\left(l-\frac{q\Phi}{2\pi}\right)\sqrt{\left[\frac{b^2m^2}{2}\left(2\left|l-\frac{q\Phi}{2\pi}\right|+3\right)\right]^{2/3}-b^2}
+2\left(2+\left|l-\frac{q\Phi}{2\pi}\right|\right)\nonumber \\
&\times&\left[\frac{b^2m^2}{2}\left(2\left|l-\frac{q\Phi}{2\pi}\right|+3\right)\right]^{1/3}-\frac{b^2m^2}{[(b^2m^2/2)(2|l-q\Phi/2\pi|+3)]^{2/3}}\Bigg]^{1/2},
\end{eqnarray}
which represents the allowed values of relativistic energy for the radial mode $n=1$ of a position-dependent mass particle subject to a linear central potential in a possible scenario described by a KKT. By comparing the Eq. (\ref{eq29}) with the result obtained in Refs. \cite{kk, kk5}, we can note that the presence of the linear potential modifies the relativistic energy levels of the system. This modification can be explained by the radical breaking of degeneracy of the Landau levels and by the representation of the lowest energy state that is defined by the radial mode $n=1$, instead of $n=0$. Again, we can note that the allowed values of relativistic energy (\ref{eq25}) are influenced by the quantum flux $\Phi$ through the shift in the eigenvalues of the angular momentum $l_{\text{eff}}=\left(l-\frac{q\Phi}{2\pi}\right)$, that is, an effect analogous to the Aharonov-Bohm effect for bound states \cite{ab, ab1}, making them a periodic function with periodicity $\Phi_0=\pm\frac{2\pi}{q}\nu$, with $\nu=0,1,2,\ldots$, that is, $\mathcal{E}_{l,1}(\Phi+\Phi_0)=\mathcal{E}_{l\mp\nu,1}(\Phi)$. This latter characteristic may be of interest in persistent current calculations \cite{me, smpr8, cp}. The result given in Eq. (\ref{eq29}), less than the term $q^2$ that stems from the KKT, is analogous to the result obtained in Ref. \cite{me1}. However, it is noteworthy that, in the latter case, the Landau gauge is inserted into the Klein-Gordon equation by the minimum coupling.

\section{Conclusion}\label{sec3}

We have investigated a position-dependent mass scalar particle subjected to a uniform magnetic field and a quantum flux in a background governed by the KKT. In this scenario, we have analyzed the interaction between a scalar particle and the Cornell-type potential, where, in the search for solutions of bound states, we determine analytically the energy profile of the system, which is influenced by the background. We can note that relativistic energy spectrum is not defined by a closed expression. In fact, it is only possible to determine allowed values of relativistic energy for each radial mode separately. As an example, we analyze the lowest energy state of the system represented by the radial mode $n=1$, instead of $n=0$. We can also note that the presence of the Cornell-type potential breaks the degeneracy of the Landau levels. In addition, a quantum effect characterized by the dependence of the magnetic field on the quantum numbers of the system is observed, where we have shown that their possible values are determined by a third-degree algebraic equation.

We have particularized our system through the parameters that characterize the linear term and the Coulomb type term of the Cornell-type potential, where we consider the absence of one or other. First, we consider the absence of the linear central potential, where we determine the allowed values of the magnetic field and the relativistic energy for the radial mode $n=1$. We can observe that both are influenced by the quantum flux through a shift in the angular momentum eigenvalues, producing an analogous effect to the Aharonov-Bohm effect for bound states, making them as periodic functions of the quantum flux. In addition, we can also note that the presence of the linear central potential breaks the degeneracy of the relativistic Landau levels. Then, we consider the absence of the linear central potential and analyze its energy profile, where, except for the expressions of the allowed values of the magnetic field and the relativistic energy for the lower energy state, which are totally modified, the characteristics of the previous case are analogous.

\acknowledgments{The authors would like to thank CNPq (Conselho Nacional de Desenvolvimento Cient\'ifico e Tecnol\'ogico - Brazil). Ricardo L. L. Vit\'oria was supported by the CNPq project No. 150538/2018-9.}


\begin{thebibliography}{99}

\bibitem{eisberg} R. Eisberg, R. Resnick, {\it{Quantum Physics of Atoms, Molecules, Solids, Nuclei and Particles}}, 2rd edn. (John Wiley, New York, 1985)

\bibitem{landau} L. D. Landau and E. M. Lifshitz, {\it{Quantum Mechanics: The Nonrelativistic Theory}}, 3rd edn. (Pergamon, Oxford, 1977)

\bibitem{smp6} A. D. Alhaidari, Phys. Rev. A {\bf{66}}, 042116 (2002)

\bibitem{smp7} A. D. Alhaidari, Int. J. Theor. Phys. {\bf{42}}, 2999 (2003)

\bibitem{smp8} A. D. Alhaidari, Phys. Lett. A {\bf{322}}, 72 (2004)

\bibitem{smp9} G. Chen, Phys. Lett. A {\bf{329}}, 22 (2004)

\bibitem{smp10} S. Dong, J. J. Pe\~na, C. Pacheco-Garc\'ía, J. Gars\'ia-Ravelo, Mod. Phys. Lett. A {\bf{22}}, 1039 (2007)

\bibitem{smp11} O. Von Roos, Phys. Rev. B {\bf{27}}, 7547 (1983)

\bibitem{smp12} O. Von Roos, H. Mavromatis, Phys. Rev. B {\bf{31}}, 2294 (1985)

\bibitem{smp13} R. A. Morrow, Phys. Rev. B {\bf{35}}, 8074 (1987)

\bibitem{smp14} W. Trzeciakowski, Phys. Rev. B {\bf{38}}, 4322 (1988)

\bibitem{smp15} I. Galbraith, G. Duggan, Phys. Rev. B {\bf{38}}, 10057 (1988)

\bibitem{smp16} K. Young, Phys. Rev. B {\bf{39}}, 13434 (1989)

\bibitem{smp17} G. T. Einevoll, P. C. Hemmer, J. Thomsen, Phys. Rev. B {\bf{42}}, 3485 (1990)

\bibitem{smp18} J. M. L\'evy Leblond, Phys. Rev. A {\bf{52}}, 1845 (1995)

\bibitem{smp19} B. G\"onul, O. \"Ozer, B. G\"on\"ul, F. \"Uzg\"un, Mod. Phys. Lett. A {\bf{17}}, 2453 (2002)

\bibitem{smp20} B. G\"on\"ul, B. G\"on\"ul, D. Tutcu and O. \"Ozer, Mod. Phys. Lett. A {\bf{17}}, 2057 (2002)

\bibitem{smp21} A. R. Plastino, M. Casas, A. Plastino, Phys. Lett. A {\bf{281}}, 297 (2001)

\bibitem{smp} T. Gora, F. Williams, Phys. Rev. {\bf{177}}, 1179 (1969)

\bibitem{bastard} G. Bastard, {\it{Wave Mechanics Applied to Semiconductor Heterostructure}} (Les Editions de Physique, France, 1988)

\bibitem{harr} P. Harrison, {\it{Quantum Wells, Wires and Dots: Theoretical and Computational Physics}}, 2rd edn. (John Wiley, UK, 2005)

\bibitem{smp1} J. Yu, S. H. Dong, G. H. Sun, Phys. Lett. A {\bf{322}}, 290 (2004)

\bibitem{smp2} J. Yu, S. H. Dong, Phys. Lett. A {\bf{325}}, 194 (2004)

\bibitem{smp3} M. Lozada-Cassou, S. H. Dong, J. Yu, Phys. Lett. A {\bf{331}}, 45 (2004)

\bibitem{smp4} F. Arias de Saavedra, J. Boronat, A. Polls, A. Fabrocini, Phys. Rev. B {\bf{50}}, 4248 (1994)

\bibitem{smp5} M. Barranco, M. Pi, S. M. Gatica, E. S. Hernandez, J. Navarro, Phys. Rev. B {\bf{56}}, 8997 (1997)

\bibitem{smp22} V. Aguiar, I. Guedes, Rev. Bras. Ens. Fís. {\bf{35}}, 1 (2013)

\bibitem{smp23} A. P. Lima, D. X. Macedo, I. Guedes, Rev. Bras. Ens. Fís. {\bf{36}}, 2305 (2014)

\bibitem{smp24} J. P. G. Nascimento, I. Guedes, Rev. Bras. Ens. Fís. {\bf{36}}, 4308 (2014)

\bibitem{greiner} W. Greiner, {\it{Relativistic Quantum Mechanics: Wave Equations}}, 3rd edition (Springer, Berlin, 2000)

\bibitem{smpr} H. G. Dosch, J. H. D. Jensen, and V. F. M\"uller, Phys. Norwegica {\bf{5}}, 2 (1971)

\bibitem{smpr1} G. Soff, B. M\"uller, J. Rafelski, W. Greiner, Z. Naturforsch {\bf{28a}}, 1389 (1973)

\bibitem{smpr2} M. K. Bahar, F. Yasuk, Adv. High Energy Phys. {\bf{2013}}, 814985 (2013)

\bibitem{smpr3} S. M. Ikhdair, M. Hamzavi, Chin. Phys. B {\bf{21}}, 110302 (2012)

\bibitem{smpr4} H. Motavalli, A. R. Akbarieh, Mod. Phys. Lett. A {\bf{25}}, 2523 (2010)

\bibitem{eug} E. R. Figueiredo Medeiros, E. R. Bezerra de Mello, Eur. Phys. J. C {\bf{72}}, 2051 (2012)

\bibitem{smpr5} L. C. N. Santos, C. C. Barros Jr., Eur. Phys. J. C {\bf{78}}, 13 (2018)

\bibitem{eug1} A. L. Cavalcanti de Oliveira, E. R. Bezerra de Mello, Class. Quantum Grav. {\bf{23}}, 5249 (2006)

\bibitem{smpr6} M. S. Cunha, C. R. Muniz, H. R. Christiansen, V. B. Bezerra, Eur. Phys. J. C {\bf{76}}, 512 (2016)

\bibitem{me} R. L. L. Vit\'oria, K. Bakke, Gen. Relativ. Gravit. {\bf{48}}, 161 (2016)

\bibitem{me1} R. L. L. Vit\'oria, K. Bakke, Int. J. Mod. Phys. D {\bf{27}}, 1850005 (2018)

\bibitem{me2} R. L. L. Vit\'oria, K. Bakke, Eur. Phys. J. Plus {\bf{133}}, 490 (2018)

\bibitem{bb} K. Bakke, H. Belich, Ann. Phys. {\bf{360}}, 596 (2015)

\bibitem{me3} R. L. L. Vit\'oria, H. Belich, K. Bakke, Adv. High Energy Phys. {\bf{2017}}, 6893084 (2017)

\bibitem{me4} R. L. L. Vit\'oria, H. Belich, Adv. High Energy Phys. {\bf{2019}}, 1248393 (2019)

\bibitem{bf} K. Bakke, C. Furtado, Ann. Phys. {\bf{355}}, 48 (2015)

\bibitem{me5} R. L. L. Vit\'oria, K. Bakke, Eur. Phys. J. Plus {\bf{131}}, 36 (2016)

\bibitem{me6} R. L. L. Vit\'oria, C. Furtado, K. Bakke, Ann. Phys. {\bf{370}}, 128 (2016)

\bibitem{me7} R. L. L. Vit\'oria, H. Belich, Eur. Phys. J. C {\bf{78}}, 999 (2018)

\bibitem{rb} R. F. Ribeiro, K. Bakke, Ann. Phys. {\bf{385}}, 36 (2017)

\bibitem{smpr7} Z. Wang, Z. Long, C. Long, M. Wu, Eur. Phys. J. Plus {\bf{130}}, 36 (2015)

\bibitem{me8} R. L. L. Vit\'oria, C. Furtado, K. Bakke, Eur. Phys. J. C {\bf{78}}, 44 (2018)

\bibitem{smpr8} E. V. B. Leite, H. Belich, K. Bakke, Adv. High Energy Phys. {\bf{2015}}, 925846 (2015)

\bibitem{kaluza} Th. Kaluza, Sitzungsber. K. Preuss. Akad. Wiss. {\bf{K1}}, 966 (1921)

\bibitem{klein} O. Z. Klein, Phys. Z. {\bf{37}}, 895 (1926)

\bibitem{kk} C. Furtado, F. Moraes, V. B. Bezerra, Phys. Rev. D {\bf{59}}, 107504 (1999)

\bibitem{kk1} V. B. Bezerra, C. Furtado, F. Moraes, Mod. Phys. Lett. A {\bf{15}}, 253 (2000)

\bibitem{kk2} G. A. Marques, C. Furtado, V. B. Bezerra, F. Moraes, J. Phys. A: Math. Gen. {\bf{34}} 5945 (2001)

\bibitem{kk3} J. G. de Assis, C. Furtado, F. Moraes, V. B. Bezerra, Gravitation and Cosmology, {\bf{6}}, 233 (2000)

\bibitem{kk4} K. Bakke, A. Yu. Petrov, C. Furtado, Ann. Phys. {\bf{327}}, 2946 (2012)

\bibitem{kk5} J. Carvalho, A. M. M. Carvalho, E. Cavalcante, C. Furtado, Eur. Phys. J. C {\bf{76}}, 365 (2016)

\bibitem{kk6} A. Macias, H. Dehnen, Class. Quantum Grav. {\bf{8}}, 203 (1991)

\bibitem{ab} Y. Aharonov, D. Bohm, Phys. Rev. {\bf{115}} 485 (1959)

\bibitem{ab1} M. Peshkin, A. Tonomura, {\it{The Aharonov-Bohm Effect in Lecture Notes in Physics}}, Vol. 340 (Springer-Verlag, Berlin, 1989)

\bibitem{fn} C. Alexandrou, P. de Forcrand, O. Jahn, Nuclear Phys. B {\bf{119}}, 667 (2003)

\bibitem{heun} A. Ronveaux, {\it{Heun's Differential Equations}} (Oxford University Press, Oxford, 1995)

\bibitem{arf} G. B. Arfken, H. J. Weber, {\it{Mathematical Methods for Physicists}}, sixth edition (Elsevier Academic Press, New York, 2005)

\bibitem{pc} H. Asada, T. Futamase, Phys. Rev. D {\bf{56}}, R6062 (1997)

\bibitem{pc1} C. L. Chrichfield, J. Math. Phys. {\bf{17}}, 261 (1976)

\bibitem{pc2} K. Bakke, Ann. Phys. {\bf{341}}, 86 (2014)

\bibitem{pc3} I. C. Fonseca, K. Bakke, J. Math. Phys. {\bf{56}}, 062107 (2014)

\bibitem{pc4} P. M. T. Barboza, K. Bakke, Ann. Phys. {\bf{361}}, 259 (2015)

\bibitem{pc5} S. M. Ikhdair, B. J. Falaye, M. Hamzavi, Ann. Phys. {\bf{353}}, 282 (2015)

\bibitem{pc6} I. I. Guseinov, B. A. Mamedov, J. Chem. Phys. {\bf{121}}, 1649 (2004)

\bibitem{pc7} I. I. Guseinov, J. Chem. Phys. {\bf{120}}, 9454 (2004)

\bibitem{pc8} A. Souza Dutra, Phys. Rev. A {\bf{47}}, R2435 (1993)

\bibitem{pc9} S. M. Ikhdair, M. Hamzavi, Physica B {\bf{407}}, 419 (2012)

\bibitem{cp} K. Bakke, C. Furtado, Ann. Phys. {\bf{336}}, 489 (2013)

\bibitem{pl} E. J. Austin, Mol. Phys. {\bf{40}}, 893 (1980)

\bibitem{pl1} E. R. Vrscay, Phys. Rev. A {\bf{31}}, 2054 (1985)

\bibitem{pl2} K. Killingbeck, Rep. Prog. Phys. {\bf{40}}, 963 (1977)

\bibitem{pl3} K. Killingbeck, Phys. Lett. A 65, 87 (1978).

\bibitem{pl4} R. P. Saxena, V. S. Varma, J. Phys. A: Math. Gen. {\bf{15}}, L149 (1982)

\bibitem{pl5} R. L. Gibbs, Am. J. Phys. {\bf{43}}, 25 (1975)

\bibitem{pl6} R. D. Desko, D. J. Bord, Am. J. Phys. {\bf{51}}, 82 (1983)

\bibitem{ball} L. E. Ballentine, {\it{Quantum Mechanics, a Modern Development}} (World Scientific, Singapore, 1998)




\end{thebibliography}
\end{document}